\documentclass[aps,prd,amssymb,cite,
amsfonts,epsf,preprintnumbers,nofootinbib,superscriptaddress]{revtex4-2}

\usepackage{graphicx}
\usepackage{bm,latexsym,amsmath,amssymb,amsfonts}
\usepackage[usenames,dvipsnames]{color}
\usepackage[colorlinks=true,linkcolor=blue]{hyperref}
\usepackage{color}
\usepackage{ulem}
\usepackage{epsfig}
\usepackage{braket}
\usepackage[mathscr]{eucal}
\usepackage{cancel}
\usepackage{mathrsfs}
\usepackage{pgf, tikz}
\usetikzlibrary{arrows.meta,positioning,shapes.geometric,calc}
\usepackage{slashed}
\usetikzlibrary{arrows,automata}

\definecolor{mypink1}{rgb}{0.858, 0.188, 0.478}
\definecolor{mypink2}{RGB}{219, 48, 122}
\definecolor{mypink3}{cmyk}{0, 0.7808, 0.4429, 0.1412}
\definecolor{mygray}{gray}{0.6}
\definecolor{pptbg}{rgb}{0.961,0.945,0.863}

\newcommand{\half}{\frac12}

\newcommand{\be}[1]{\begin{equation} \label{#1}}
\newcommand{\ee}{\end{equation}}
\newcommand{\bea}{\begin{eqnarray}}
\newcommand{\eea}{\end{eqnarray}}
\newcommand{\ba}{\begin{array}}
\newcommand{\ea}{\end{array}}
\newcommand{\nn}{\nonumber}

\newcommand{\bel}{\begin{align}}
\newcommand{\eel}{\end{align}}

\newcommand{\Lie}{\pounds}

\newcommand{\hreff}[1]{\href{#1}{\color{blue}{#1}} }
\renewcommand{\d}{\mathrm{d}}

\begin{document}
\title{Particle creation, clock variables, and relabelling symmetry in pull-back variational fluids}

\author{Hyeong-Chan Kim}
\affiliation{School of Liberal Arts and Sciences, Korea National University of Transportation, Chungju 380-702, Korea}

\email{hckim@ut.ac.kr}


\begin{abstract}
Particle-number conservation is usually built into the pull-back variational formulation of relativistic fluids. The standard reason is that the conservative Eulerian variation of the number current is compatible with flow-aligned relabellings only when \(\nabla_a n^a=0\). We revisit this argument when the matter-space density itself carries particle-creation information. The resulting structure is best organized by the fate of the flow-aligned relabelling. If particle creation is encoded by the proper-time functional of the deformed worldline, the endpoint variation of the proper time cancels the timelike creation residual, so that the relabelling can still be treated as a current-level gauge symmetry. The price is a history term proportional to the acceleration, which disappears on geodesic flows, including homogeneous FLRW cosmological backgrounds. By contrast, local descriptions based on an interacting matter space or on an independent scalar clock do not in general preserve the current-level relabelling symmetry when \(\Gamma_N\neq 0\). Instead, the Eulerian current variation contains a timelike residual proportional to \(\Gamma_Nu^a\). In the multi-fluid construction of Andersson and Comer, where the density three-form of one species may depend on the matter-space coordinates of another, this residual is absorbed into the coupled force balance and describes inter-species conversion within the variational principle. In single-flow self-creation, no partner matter-space force is available, and the same residual becomes the physical scalar mode carried by the creation clock. When the clock is a standard local scalar satisfying \(u^a\nabla_a\Theta=1\), its dynamics or constraint supplies the scalar equation replacing particle-number conservation. In a dissipative system this scalar clock may be interpreted as an entropic time, a local clock associated with the thermodynamic arrow selected by entropy production. 
Although these clock choices can give the same homogeneous FLRW creation history, their linear perturbations need not agree: the proper-time clock is locked to the perturbed expansion, whereas a local scalar clock generically carries an independent entropy mode that contributes to the creation-pressure perturbation and to the effective sound speed.
\end{abstract}
\maketitle
 
\section{Introduction}

In high-temperature and dissipative systems, particle creation and destruction are not exceptional processes; see, for example, Ref.~\cite{Ichiyanagi1994} for a review of variational approaches to irreversible processes.
Chemical reactions may continuously change the number of particles in each species, and the associated change of composition is inseparable from heat production in non-equilibrium thermodynamics. 
In cosmology, particle creation in the early universe has long been studied both microscopically in quantum field theory in curved spacetime~\cite{Parker1969,BirrellDavies}
and phenomenologically in open-system cosmology~\cite{Prigogine1988,Lima1996}.
From this point of view, particle-number conservation is not a first-principles law on the same footing as energy-momentum conservation. 
It is rather a simplifying assumption whose validity depends on the physical regime under consideration.

Nevertheless, particle-number conservation is built deeply into many variational formulations of relativistic fluids, especially in the Taub--Carter pull-back approach~\cite{Taub54,Carter72,Carter73,Carter89,Lopez2011,AnderssonNew}.
In the pull-back construction, the number current is obtained from a three-form on matter space. If this matter-space three-form depends only on the matter-space coordinates of the same species, its pull-back is closed by construction, and the corresponding number current is conserved. The same conclusion also appears from the variational point of view. The conservative Eulerian variation of the number current is compatible with a flow-aligned relabelling of the fluid elements only when
\[
\Gamma_N\equiv\nabla_a n^a=0 .
\]
This observation is often interpreted as a variational obstruction to particle creation.

The purpose of this paper is to revisit this obstruction. We will argue that the obstruction is not a universal prohibition of particle creation in the pull-back formalism. Rather, it reflects how particle creation is encoded in the matter-space density and how the flow-aligned relabelling is interpreted. The resulting structure is best organized by the fate of the flow-aligned relabelling, and three physically distinct realizations must be separated.

The first realization preserves the relabelling as a current-level gauge symmetry. This is achieved by identifying the creation clock with the proper-time functional of the deformed worldline. The endpoint variation of the proper time then cancels the timelike creation residual, so that the flow-aligned relabelling can still be treated as a gauge direction even when \(\Gamma_N\neq0\). The price is a history term proportional to the acceleration of the flow, which vanishes on geodesic congruences, including homogeneous FLRW cosmological backgrounds.

The remaining realizations are local descriptions in which the relabelling is no longer a current-level gauge symmetry. The Eulerian current variation then retains a purely timelike residual proportional to \(\Gamma_Nu^a\), and the question becomes how this residual is represented in the variational principle.

The second realization is self-creation by a single species, in which the particle number of one species changes without being represented as conversion from another matter-space current. Here there is no partner matter space and no interaction term that can absorb the residual. We introduce a local scalar clock \(\Theta\) that carries the information about particle creation along the worldline. The residual then survives as the physical scalar mode of the clock, which in a dissipative system may be interpreted as a local entropic time selected by entropy production.

The third realization is inter-species conversion. In the multi-fluid construction of Andersson and Comer~\cite{Andersson:2013jga,AnderssonNew}, the matter-space density three-form of one species may depend on the matter-space coordinates of another species. This dependence produces a local interaction term in the variation. The Eulerian current variation of a single species still contains the timelike residual, but the residual is now absorbed by the partner-sector interaction force in the action variation, so that the flow-aligned direction drops out of the coupled force balance. Physically, the creation of one species is then tied to conversion from, or into, another species, and the partner matter-space fields carry the compensating variation.

The central result is that particle creation is compatible with pull-back variational fluid dynamics, but it reorganizes the status of the flow-aligned relabelling: a proper-time clock preserves the relabelling at the price of a possible history term, whereas local descriptions expose a timelike residual that is either carried by a physical clock mode (self-creation) or absorbed by an interacting partner sector (inter-species conversion).
The three possibilities are summarized schematically in
Fig.~\ref{fig:relabelling-flowchart}.

\begin{figure}[t]
\centering
\begin{tikzpicture}[
    node distance=1.4cm and 1.9cm,
    box/.style={
        rectangle,
        rounded corners,
        draw=black,
        align=center,
        text width=3.4cm,
        minimum height=0.95cm,
        font=\small
    },
    widebox/.style={
        rectangle,
        rounded corners,
        draw=black,
        align=center,
        text width=5.0cm,
        minimum height=0.95cm,
        font=\small
    },
    smallbox/.style={
        rectangle,
        rounded corners,
        draw=black,
        align=center,
        text width=3.2cm,
        minimum height=0.85cm,
        font=\small
    },
    arrow/.style={
        -{Latex[length=2.0mm]},
        thick
    }
]

\node[widebox] (start)
{Pull-back fluid \\ with particle creation};

\node[widebox, below=of start] (residual)
{Flow-aligned relabelling residual\\
\(\sim \Gamma_N u^a\)};

\node[box, below left=1.5cm and 2.0cm of residual] (proper)
{Proper-time clock\\
\(n_{ABC}(N,\tau)\)};

\node[box, below=1.5cm of residual] (local)
{Local scalar clock\\
\(n_{ABC}(N,\Theta)\)};

\node[box, below right=1.5cm and 3.0cm of residual] (multi)
{Inter-species conversion\\
\(n_{ABC}(N,S)\)};

\node[smallbox, below=of proper] (proper2)
{Endpoint variation\\
cancels residual};

\node[smallbox, below=of local] (local2)
{Residual remains\\
at current level};

\node[smallbox, below=of multi] (multi2)
{No separate clock\\
variable};

\node[smallbox, below=of proper2] (proper3)
{Current-level gauge\\
symmetry preserved};

\node[smallbox, below=of local2] (local3)
{Physical scalar\\
clock/entropic mode};

\node[smallbox, below=of multi2] (multi3)
{Partner matter space\\
carries conversion};

\node[smallbox, below=of proper3] (proper4)
{\(\Delta_\Gamma=0\)\\
in minimal FLRW};

\node[smallbox, below=of local3] (local4)
{\(\Delta_\Gamma\neq0\)\\
generically};

\node[smallbox, below=of multi3] (multi4)
{Residual absorbed by\\
relative force balance \(n^a{\cal E}^N_a=0\)\\
};

\draw[arrow] (start) -- (residual);

\draw[arrow] (residual.south) -- ++(0,-0.55) -| (proper.north);
\draw[arrow] (residual.south) -- (local.north);
\draw[arrow] (residual.south) -- ++(0,-0.55) -| (multi.north);

\draw[arrow] (proper) -- (proper2);
\draw[arrow] (proper2) -- (proper3);
\draw[arrow] (proper3) -- (proper4);

\draw[arrow] (local) -- (local2);
\draw[arrow] (local2) -- (local3);
\draw[arrow] (local3) -- (local4);

\draw[arrow] (multi) -- (multi2);
\draw[arrow] (multi2) -- (multi3);
\draw[arrow] (multi3) -- (multi4);

\end{tikzpicture}
\caption{
Fate of the flow-aligned relabelling residual in the three realizations
of particle creation considered in this work.  A proper-time clock cancels
the timelike residual at the level of the current variation and preserves
the relabelling as a gauge direction.  A standard local scalar clock leaves
the residual as a physical scalar creation mode.  In the Andersson--Comer
inter-species construction, the residual remains at the current level but
is absorbed by the coupled force balance through the partner-sector
interaction force.
}
\label{fig:relabelling-flowchart}
\end{figure}

The paper is organized as follows. In Sec.~\ref{sec:obstruction}, we review the conservative pull-back construction and the standard relabelling obstruction that leads to \(\Gamma_N=0\). In Sec.~\ref{sec:propertime}, we show that identifying the creation clock with the proper-time functional preserves the flow-aligned relabelling as a gauge symmetry, at the price of a history term proportional to the acceleration. In Sec.~\ref{sec:self}, we treat the clock as an ordinary local scalar field and show that the flow-aligned relabelling then exposes a timelike residual, which for self-creation must be carried as a physical scalar mode and may be interpreted as an entropic time. In Sec.~\ref{sec:AC}, we show how the Andersson--Comer multi-fluid construction realizes the same residual at the current level but absorbs it into the coupled force balance for inter-species conversion. 
In Sec.~\ref{sec:cosmol}, we illustrate the construction on an FLRW background.  We also show that the equivalence of the proper-time and local scalar-clock descriptions at the homogeneous level is broken by perturbations: the former gives a creation-rate perturbation fixed by the expansion perturbation, whereas the latter contains an independent clock mode and therefore an intrinsic non-adiabatic creation-pressure perturbation.
Section~\ref{sec:summary} summarizes the results and discusses the interpretation of the relabelling residual.

\section{Relabelling obstruction in the conservative pull-back formalism}
\label{sec:obstruction}

We begin by recalling why particle-number conservation appears naturally in the standard pull-back variational formulation. 
Consider a single number current \(n^a\). Following the pull-back construction of Carter and collaborators~\cite{Carter72,Carter73,Carter89,AnderssonNew}, it is convenient to introduce a three-dimensional matter space 
with coordinates \(N^A\), \(A=1,2,3\). Each scalar field \(N^A(x)\) labels a fluid element, and the labels are constant along the corresponding flow line. A matter-space three-form
\[
\bm{n}=n_{ABC}(N^D)\,dN^A\wedge dN^B\wedge dN^C
\]
is pulled back to spacetime as
\be{n st}
n_{abc}
= n_{ABC}
\frac{\partial N^A}{\partial x^a}
\frac{\partial N^B}{\partial x^b}
\frac{\partial N^C}{\partial x^c} .
\ee
The number current is then defined by
\be{n}
n^a\equiv \frac{1}{3!}\epsilon^{abcd}n_{bcd}.
\ee
When \(n_{ABC}\) depends only on the matter-space coordinates \(N^D\), the pulled-back three-form is closed by construction. Equivalently,
\[
\nabla_{[a}n_{bcd]}=0,
\qquad
\nabla_a n^a=0 .
\]
Thus particle-number conservation is built into the conservative pull-back construction.

The same restriction can be seen from the variational relation. Let \(\xi^a\) be the Lagrangian displacement of the fluid element. The conservative Eulerian variation of the number current is~\cite{Carter72,Carter73,Carter89,AnderssonNew} 
\be{d na}
\delta n^a
= -\Lie_\xi n^a
-n^a\left(\nabla_b\xi^b+\frac12 g^{bc}\delta g_{bc}\right),
\ee
where \(\Lie_\xi\) denotes the Lie derivative along \(\xi^a\). The unit velocity and the number density are defined by
\be{u n}
u^a\equiv \frac{n^a}{n},
\qquad
n\equiv \sqrt{-n^a n_a}.
\ee

The pull-back description contains a relabelling freedom. Sliding the labels of a fluid element along the same physical worldline should not change the physical configuration. Therefore, two displacements \(\xi^a\) and
\[
\bar\xi^a=\xi^a-G^a
\]
should generate equivalent variations when \(G^a\) is parallel to the current. Using Eq.~\eqref{d na}, the difference between the two variations is
\bea
(\delta-\bar\delta)n^a
&=&
2\nabla_b\left(n^{[b}G^{a]}\right)
-G^a\nabla_b n^b .
\label{proof1}
\eea
where $\bar \delta$ denotes the variation generated by $\bar \xi^a$. 
For the proof, see Sec.~6 of Ref.~\cite{AnderssonNew}. 
If \(G^a\) is flow-aligned, \(G^a=G u^a\), the antisymmetric term vanishes. The remaining term is
\[
(\delta-\bar\delta)n^a=-G u^a \Gamma_N,
\qquad
\Gamma_N\equiv\nabla_a n^a .
\]
Hence the two variations are equivalent only if
\be{GN=0}
\Gamma_N \equiv \nabla_a n^a=0 .
\ee
This is the standard relabelling obstruction. In the conservative pull-back formalism, flow-aligned relabelling is a gauge symmetry only for a conserved number current.

This result has an important conceptual consequence. In a conservative fluid, sliding a label along a worldline does not change the particle content assigned to that worldline. The corresponding direction in the space of variations is therefore redundant. The condition~\eqref{GN=0} is then both a conservation law and the condition that removes the flow-aligned relabelling direction from the physical variations.

However, this argument relies on the conservative variation~\eqref{d na}. It assumes that the matter-space density contains no additional information about particle creation. Once the density three-form is allowed to depend either on a clock variable or on another matter space, the Eulerian variation acquires additional terms. 
The conclusion \(\Gamma_N=0\) is then no longer automatic. 
However, the additional terms do not necessarily restore current-level relabelling invariance. 
They determine how the timelike residual is represented in the variational principle.
In the next section we show how this plays out when the creation information is carried by a proper-time clock, where the endpoint variation of the proper time restores the relabelling as a gauge symmetry.

\section{Gauge-preserving particle creation: the proper-time clock}
\label{sec:propertime}

In the conservative formalism of Sec.~\ref{sec:obstruction}, the flow-aligned relabelling is a gauge symmetry precisely because the number current is conserved. We now ask how this structure changes once particle creation is encoded in the matter-space density of a single species. By self-creation we mean a process in which the particle number of one species changes without being represented as conversion from a second matter-space current. Such a situation may arise, for example, in an effective description of particle production by a background field, or in a system where the microscopic process is not resolved into separate reacting species. A new scalar datum is then needed to tell how the particle content assigned to each worldline changes; we call this datum the creation clock.

In this section we examine the simplest choice, in which the clock is identified with the proper time along the fluid worldline. This possibility was already considered in Carter's construction~\cite{Carter72,Carter73}.
As we show, this choice preserves the flow-aligned relabelling as a current-level gauge symmetry even when \(\Gamma_N\neq0\), at the price of a history term in the variation.

The matter space \(N^A\), \(A=1,2,3\), is constructed as before, but the matter-space density is now allowed to depend on a proper-time variable \(\tau\):
\[
n_{ABC}=n_{ABC}(N^D,\tau),
\qquad
u^a\nabla_a\tau=1 .
\]
Here \(\tau\) should not be understood as an independent ordinary scalar field; it is the clock reading obtained by integrating proper time along the deformed fluid worldline.
The spacetime three-form is obtained by the pull-back
\[
n_{abc}(x)
=
\frac{\partial N^A}{\partial x^{[a}}
\frac{\partial N^B}{\partial x^b}
\frac{\partial N^C}{\partial x^{c]}}
n_{ABC}(N^D,\tau).
\]
Since \(n_{ABC}\) now depends on \(\tau\), the pulled-back three-form is no longer closed. Its exterior derivative gives
\bea
\nabla_{[a}n_{bcd]}
&=&
\frac{\partial \tau}{\partial x^{[a}}
\frac{\partial N^B}{\partial x^b}
\frac{\partial N^C}{\partial x^c}
\frac{\partial N^D}{\partial x^{d]}}
\left(\frac{\partial n_{BCD}}{\partial \tau}\right).
\eea
Thus the particle creation rate is
\be{GN}
\Gamma_N\equiv\nabla_a n^a
=(\nabla_a\tau)\tilde n^a,
\qquad
\tilde n^a\equiv
\frac1{3!}\epsilon^{abcd}
\frac{\partial N^B}{\partial x^b}
\frac{\partial N^C}{\partial x^c}
\frac{\partial N^D}{\partial x^d}
\left(\frac{\partial n_{BCD}}{\partial \tau}\right).
\ee
The vector \(\tilde n^a\) measures the change of the matter-space three-form with respect to the clock variable. Since \(\partial_\tau n_{BCD}\) is again a three-form on the same matter space, \(\tilde n^a\) is parallel to \(n^a\), and hence to \(u^a\).

We next compute the variation of the number current. The variation of \(n^a\) contains the usual conservative pull-back part and an additional contribution from the clock dependence. One obtains
\bea
\delta n^a
&=&
-\Lie_\xi n^a
- n^a\left(
\nabla_b\xi^b+\frac12 g^{bc}\delta g_{bc}
\right)
+ \tilde n^a
\left[ \int^\tau (a_e\xi^e)\,\d\tau' - \xi^e u_e
\right].
\label{delta n0}
\eea
Here \(a^b\equiv u^c\nabla_cu^b\) is the acceleration of the flow. The origin of the last term is the variation of the proper-time functional. If
\[
\tau(\lambda)=\int^\lambda \d\lambda' \,
\mathcal{L}(\lambda'),
\qquad
\mathcal{L}\equiv\sqrt{-g_{ab}\dot x^a\dot x^b},
\]
then, neglecting the metric variation in the present displacement analysis,
\bea
\delta\tau
&=&
\int^\lambda \d\lambda'
\left[
\frac{\partial\mathcal{L}}{\partial x^a}
-
\frac{\d}{\d\lambda'}
\frac{\partial\mathcal{L}}{\partial\dot x^a}
\right]\delta x^a
+
\left[
\frac{\partial\mathcal{L}}{\partial\dot x^a}\delta x^a
\right]^\lambda
\nn\\
&=&
\int^\tau (a_e\xi^e)\, \d\tau'
-\xi^e u_e .
\label{delta-tau}
\eea
Substituting this result into the variation of the pulled-back three-form gives Eq.~\eqref{delta n0}.

\subsection{Meaning of the proper-time-clock variation}
\label{subsec:meaning-propertime-clock}

Equation~\eqref{delta n0} is the key result of the proper-time-clock construction. It shows that particle creation modifies the conservative variation in a very specific way. The first two terms are the standard pull-back variation of a conserved current. The last term is new. It is proportional to \(\tilde n^a\), and therefore points along the flow direction. This is expected: changing the clock changes the amount of matter assigned to the same worldline, not the spatial direction of the current.

There are two distinct contributions in the clock variation. The endpoint term,
$
-\xi^e u_e ,
$
is local. It measures how much the displacement moves the endpoint forward or backward along the fluid proper time. A flow-aligned displacement changes the clock reading directly and hence changes the amount of particle creation accumulated on the worldline.

The second contribution,
\[
\int^\tau (a_e\xi^e)\, \d\tau',
\]
is a history term. It appears because \(\tau\) is not introduced as an independent local scalar field, but as the proper-time functional of the deformed worldline. When the worldline is displaced, the accumulated proper time changes not only at the endpoint but also through the deformation of the entire trajectory. This dependence is controlled by the acceleration. Hence the non-locality in Eq.~\eqref{delta n0} is not caused by particle creation itself. It is caused by identifying the creation clock with the proper-time functional of the path.

This distinction is important. On a geodesic flow, \(a^b=0\), the history term vanishes and the clock variation becomes local:
\[
\delta\tau=-\xi^e u_e .
\]
Thus the proper-time-clock construction is local on geodesic congruences, including homogeneous FLRW cosmological backgrounds. For accelerated flows, however, the proper-time functional carries memory of the deformation of the path, and the variational relation becomes non-local.

We can now examine the relabelling issue. Consider two displacements \(\xi^a\) and $\bar\xi^a=\xi^a-G^a .$
Using Eq.~\eqref{delta n0}, the difference between the two variations is
\bea
(\delta-\bar\delta)n^a
&=&
\nabla_b(n^bG^a-n^aG^b)
-
\left[
G^a\Gamma_N
+
(G^bu_b)\tilde n^a
\right]
+
\tilde n^a\int^\tau(a_eG^e)\, \d\tau' .
\label{eq:diff-propertime-clock}
\eea
For a flow-aligned relabelling, \(G^a=Gu^a\), the divergence term vanishes. The history term also vanishes because \(a_eu^e=0\). The remaining term is
$
-G(u^a\Gamma_N-\tilde n^a).
$
Since \(\tilde n^a\) is parallel to \(u^a\), and since
\[
\Gamma_N=(\nabla_a\tau)\tilde n^a,
\qquad
u^a\nabla_a\tau=1,
\]
we obtain
\be{tilde n}
\tilde n^a=\Gamma_N u^a .
\ee
Therefore the flow-aligned relabelling residual vanishes in the proper-time-clock prescription:
\[
(\delta-\bar\delta)n^a=0 .
\]

This result has a clear interpretation. If the creation clock is taken to be the proper-time reading along the same worldline, then sliding the matter-space labels along the flow changes the endpoint clock reading in precisely the way needed to compensate the apparent creation residual. In this convention, the flow-aligned relabelling can still be treated as a gauge transformation, even when \(\Gamma_N\neq0\). The price is that the clock is a proper-time functional, and for accelerated flows its variation contains the non-local history term.

Putting Eq.~\eqref{tilde n} into Eq.~\eqref{delta n0}, the current variation becomes
\be{d n}
\delta n^a
= -\Lie_\xi n^a -n^a\left[\nabla_b\xi^b
+\xi^b n_b\frac{\Gamma_N}{n^2}-\frac{\Gamma_N}{n}
\int^\tau(a_b\xi^b)\,\d\tau' +\frac12 g^{bc}\delta g_{bc}
\right].
\ee
This formula summarizes the proper-time-clock description. It gives a variational relation compatible with non-vanishing particle creation, but it does so by tying the clock to the deformed worldline. It is therefore local only when the acceleration term vanishes.

This motivates a second, more local description. Instead of identifying the clock with the proper-time functional, one may introduce an independent scalar clock field \(\Theta\). The question then changes. One no longer asks whether the proper-time endpoint variation cancels the relabelling residual. Rather, one asks how an ordinary local scalar clock transforms under the Lagrangian displacement, and what this implies for the status of the flow-aligned relabelling.

\section{Local scalar clocks and the timelike residual}\label{sec:self}

Motivated by the non-locality of the proper-time clock, we now treat the creation clock as an ordinary local scalar field \(\Theta\). Throughout this section we use \(\chi_a\) to denote the momentum covector conjugate to the number current \(n^a\), defined through the variation \(\delta\Lambda=\chi_a\delta n^a+\dots\) of the matter Lagrangian $\Lambda$, and we write \(\chi\equiv -u^a\chi_a\) for its magnitude along the flow. We first record two possible clock conditions, then derive the variation of the number current and the fate of the flow-aligned relabelling.

\subsection{Strong and weak clock conditions}
\label{subsec:strong-weak-clock}

Before introducing an independent scalar clock, it is useful to distinguish two possible clock conditions. One may try to identify the clock with an orthogonal proper-time potential,
\be{eq:strongclock1}
\nabla_a\Theta=-u_a .
\ee
This is a strong condition. It implies the weaker condition
\be{eq:weakclock1}
u^a\nabla_a\Theta=1 
\ee
but the converse is not true. The weaker condition states only that \(\Theta\) advances as proper time along each worldline.

For a general scalar clock one may decompose
\be{eq:clockdecomp-general}
\nabla_a\Theta
=
-Cu_a+D_a\Theta,
\qquad
C\equiv u^b\nabla_b\Theta,
\qquad
D_a\Theta\equiv h_a{}^b\nabla_b\Theta,
\qquad
u^aD_a\Theta=0 .
\ee
The weak clock condition is the normalization \(C=1\). Thus, under Eq.~\eqref{eq:weakclock1},
$
\nabla_a\Theta=-u_a+D_a\Theta ,
$
and the hypersurfaces \(\Theta=\mathrm{const.}\) need not be orthogonal to the flow.

The strong condition is much more restrictive. With
$ a_b\equiv u^c\nabla_cu_b,$ $h_{ab}\equiv g_{ab}+u_au_b,$ and $
\omega_{ab}\equiv h_a{}^ch_b{}^d\nabla_{[c}u_{d]}, $
the kinematical decomposition gives
$
\nabla_a u_b
=
-u_aa_b
+
\frac13\theta h_{ab}
+
\sigma_{ab}
+
\omega_{ab},
$
and hence
\be{eq:decomp1}
2\nabla_{[a}u_{b]}
= 2\omega_{ab} -2u_{[a}a_{b]} .
\ee
If \(\nabla_a\Theta=-u_a\), then \(u_a\) is exact and
$
\nabla_{[a}u_{b]}=0 .
$
Equation~\eqref{eq:decomp1} then implies
\[
\omega_{ab}=0,
\qquad
a_b=0 .
\]
Conversely, if the flow is geodesic and irrotational, then \(du=0\), and the Poincar\'e lemma gives a local scalar \(\Theta\) satisfying Eq.~\eqref{eq:strongclock1}. Therefore the strong clock exists locally if and only if the flow is geodesic and irrotational. By contrast, the weak condition~\eqref{eq:weakclock1} is a transport equation along \(u^a\) and admits local solutions for arbitrary flows.

\subsection{A standard scalar clock}
\label{subsec:standard-scalar-clock}

The proper-time-clock construction shows that particle creation can be incorporated into the variational relation, but it also shows why a proper-time functional is not the most local description for general flows. We therefore introduce an independent scalar clock field \(\Theta\) and allow
\be{eq:nABCTheta}
n_{ABC}=n_{ABC}(N^D,\Theta).
\ee
We now treat \(\Theta\) as an ordinary scalar under the Lagrangian displacement:
\be{eq:DeltaThetaZero}
\Delta\Theta=0,
\qquad
\delta\Theta=-\xi^a\nabla_a\Theta ,
\ee
$\Delta \Theta=0$ is used here as the variational prescription for an ordinary scalar under the displacement, not as a dynamical statement that $\Theta$ is conserved along the flow.
Defining
\be{eq:ntildeTheta}
\tilde n^a
\equiv
\frac1{3!}\epsilon^{abcd}
\frac{\partial N^B}{\partial x^b}
\frac{\partial N^C}{\partial x^c}
\frac{\partial N^D}{\partial x^d}
\frac{\partial n_{BCD}}{\partial\Theta},
\ee
we again have \(\tilde n^a=\alpha u^a\), since \(\partial_\Theta n_{BCD}\) is a three-form on the same matter space. The creation rate is
\be{eq:GammaTheta}
\Gamma_N
= \nabla_an^a
= (\nabla_a\Theta)\tilde n^a
=\alpha u^a\nabla_a\Theta .
\ee
Thus, provided \(u^a\nabla_a\Theta\neq0\),
\be{eq:ntildeGamma1}
\tilde n^a
=
\frac{\Gamma_N}{u^c\nabla_c\Theta}\,u^a .
\ee
With the weak normalization~\eqref{eq:weakclock1}, this reduces to Eq.~\eqref{tilde n}.

The variation of the number current is now local:
\be{eq:dnThetaStandard}
\delta n^a
= -\Lie_\xi n^a
-n^a\left(\nabla_b\xi^b+\half g^{bc}\delta g_{bc}\right) 
 -\tilde n^a\xi^b\nabla_b\Theta .
\ee
No history integral appears. Hence particle creation does not by itself force a non-local variational relation. The non-locality in Eq.~\eqref{d n} is tied specifically to the use of the proper-time functional as the creation clock.

The price of this locality is conceptual. Because \(\Theta\) is now an ordinary scalar field, the flow-aligned relabelling no longer has to be a gauge symmetry. Indeed, for two displacements \(\xi^a\) and \(\bar\xi^a=\xi^a-G^a\), Eq.~\eqref{eq:dnThetaStandard} gives
\bea
(\delta-\bar\delta)n^a
&=&
\nabla_b(n^bG^a-n^aG^b)
- G^a\Gamma_N
-\tilde n^aG^b\nabla_b\Theta .
\label{eq:gendiffStandard}
\eea
For \(G^a=Gu^a\), the divergence term vanishes, and using Eq.~\eqref{eq:ntildeGamma1} one obtains
\be{eq:residualStandard}
(\delta-\bar\delta)n^a
=
-2G\Gamma_Nu^a = -2\Gamma_N (\xi-\bar \xi)^a.
\ee
This expression contains several pieces of information. First, the residual is proportional to the relabelling displacement \((\xi-\bar\xi)^a\), so it measures precisely the failure of the flow-aligned relabelling to be a gauge direction. Second, it is proportional to the creation rate \(\Gamma_N\), and therefore disappears in the conservative limit. Third, with the standard scalar-clock convention \(\Delta\Theta=0\), or equivalently \(\delta\Theta=-\xi^a\nabla_a\Theta\), the coefficient is fixed to be \(-2\). 
One factor of \(-1\) is the conservative relabelling residual, while the second comes from the variation of the ordinary scalar clock. Thus the factor of \(2\) is not accidental: it records the fact that the scalar-clock contribution adds to, rather than cancels, the conservative residual.

The sign should also be read with the convention used above. Since \((\xi-\bar\xi)^a=G u^a\), a positive creation rate produces a residual opposite to the relabelling displacement in the difference \((\delta-\bar\delta)n^a\). 
In this limited variational sense, the residual behaves like a response against treating the flow-aligned displacement as a gauge slide.
However, it becomes a genuine dynamical restoring force only after a clock-sector action or constraint is specified.
If one changes the clock transformation law, for example by using the clock-reading prescription \(\delta\Theta=+\xi^a\nabla_a\Theta\), this coefficient changes and may vanish. Therefore the precise numerical coefficient is convention-dependent in that broader sense, but within the standard local scalar-clock prescription the coefficient \(-2\) is meaningful. The invariant conclusion is that an ordinary local scalar clock leaves a timelike residual proportional to both \(\Gamma_N\) and the flow-aligned relabelling displacement.
This timelike structure will be useful for comparison with the inter-species case discussed in Sec.~\ref{sec:AC}. The important point at this stage is that a local scalar-clock description does not remove the flow-aligned residual at the level of the current variation.

This residual should not be interpreted as an inconsistency. It indicates that, for self-creation described by a standard local scalar clock, the flow-aligned relabelling has become a physical scalar direction. In the conservative theory, sliding the labels along a worldline does not change the physical state because the particle content on the worldline is preserved. When \(\Gamma_N\neq0\), the same slide changes the clock value and therefore the amount of matter created or destroyed. The formerly gauge direction is precisely the scalar mode that carries particle creation.

\subsection{Degree-of-freedom counting and the role of the creation clock}
\label{subsec:dofcounting}

The above discussion also clarifies the degree-of-freedom counting. 
Let us recall the situation in the standard two-fluid model for relativistic heat conduction~\cite{Monsalvo2011,Andersson2011,LK2022,Kim:2023lta,Kim:2024hls}, consisting of one particle-number current \(n^a\) and one entropy, or caloric, current \(s^a\). 
For standard approaches to relativistic dissipative thermodynamics, see
Refs.~\cite{IsraelStewart1979,Ichiyanagi1994,Monsalvo2011,Andersson2011}.
Before imposing any dynamical equations, the two currents contain eight field components. 
A closed formulation therefore requires eight independent equations of motion.

In the conservative setting these equations are supplied in the following way. The spatial projection of energy-momentum conservation,
$
\gamma_{bc}\nabla_a T^{ab}=0,
$
provides three equations.
A constitutive entropy-production relation, compatible with
$
\nabla_a s^a \geq 0 ,
$
and, in simple heat-conducting models, of the form
$
\nabla_a s^a \propto q^2 ,
$
provides one scalar closure relation.
The relativistic Cattaneo-type equation for the heat flux supplies three further equations~\cite{Monsalvo2011,Andersson2011,LK2022}.
The remaining scalar equation is normally taken to be the particle-number conservation law~\eqref{GN=0}.
In this way the eight components of \((n^a,s^a)\) are matched by eight equations.

This counting is closely tied to the relabelling structure of the pull-back formalism. For a conserved number current, the displacement along the flow direction is a redundancy. Sliding the matter-space labels along the same physical worldline does not change the particle content carried by that worldline. The flow-aligned relabelling is therefore a gauge direction, and the condition~\eqref{GN=0} serves both as the conservation law and as the scalar equation associated with the removal of this redundant direction.

The situation changes when particle creation is allowed. If \(\Gamma_N\neq0\), the particle content assigned to a given worldline changes with the creation clock. A displacement along the flow direction then changes the amount of matter created or destroyed. In the self-creation case, where there is no partner matter space to compensate this change, the flow-aligned relabelling is no longer a gauge redundancy. The scalar direction that was removed in the conservative theory becomes a physical degree of freedom.

This observation explains why one should not simply drop the equation
$\nabla_a n^a=0$ without replacing it by another relation.
Once particle-number conservation is abandoned, the conservative scalar equation is lost.
At the same time, the formerly gauge flow-aligned direction becomes physical.
Therefore the theory requires an additional scalar equation governing the creation process.
In the present formulation this equation is naturally associated with the clock variable that carries the information about particle creation along the worldline.

For a standard local scalar clock \(\Theta\), the creation rate is determined by the dependence of the matter-space density on \(\Theta\). If $n_{ABC}$ depends on $\Theta$ as in Eq.~\eqref{eq:nABCTheta}, then the creation rate becomes Eq.~\eqref{eq:GammaTheta}.
Since \(\tilde n^a\) is parallel to \(u^a\), the clock controls the scalar creation mode. If the weak clock normalization~\eqref{eq:weakclock1} is imposed, then \(\tilde n^a=\Gamma_N u^a\), the same formula in Eq.~\eqref{tilde n}. The missing scalar equation may then be supplied either by a prescribed creation law, by a constraint on \(\Theta\), or by an equation of motion obtained from an additional clock-sector action. The obstruction posed by a freely varied algebraic clock, and the explicit clock dynamics that supplies this scalar equation, are taken up in Sec.~\ref{subsec:localaction}.

In summary, the equation counting is not spoiled by particle creation; rather, it is reorganized. In the conservative theory, the flow-aligned relabelling is gauge and the scalar equation is \(\Gamma_N=0\). In the self-creating theory with a standard clock, the same direction becomes physical, and the scalar equation must be supplied by the dynamics or constraint of the clock. The clock field therefore plays a dual role: it records the change of particle content along each worldline and provides the additional scalar relation that replaces particle-number conservation.

\subsection{Entropic time as the creation clock}
\label{subsec:entropictime}

The clock variable introduced above need not be regarded as an external or purely auxiliary field. 
In a dissipative system there is a natural internal notion of time: the direction selected by entropy production~\cite{Ichiyanagi1994}. 
Since particle creation is one of the microscopic mechanisms by which a non-equilibrium system changes its composition and produces heat, it is natural to identify the creation clock with an entropic time.
This viewpoint is related in spirit to the thermal-time idea, in which a
thermodynamic state selects a time flow~\cite{ConnesRovelli1994}, although here we use only a local hydrodynamic clock associated with entropy production.

We do not assume that the clock is identical to the total entropy. The total entropy is an extensive quantity and, in a relativistic local theory, its precise definition may depend on coarse graining and on the choice of hypersurface. Instead, we introduce a local scalar field \(\Theta_{\rm ent}\) whose monotonic increase along the flow represents the thermodynamic arrow of time. A minimal requirement is
\be{eq:entropicclockmonotonic}
u^a\nabla_a\Theta_{\rm ent}>0
\ee
whenever irreversible entropy production is present. More generally, one may relate the rate of the entropic clock to the entropy production rate by
\be{eq:entropicclockrate}
u^a\nabla_a\Theta_{\rm ent}
=
\mathcal{F}(\Gamma_S,s,n,\chi,\Theta_{\rm ent},\cdots),
\qquad
\Gamma_S\equiv\nabla_a s^a ,
\ee
where \(\mathcal{F}\) is positive when \(\Gamma_S>0\). The weak clock normalization used above corresponds to choosing the parametrization of the entropic clock such that
\be{eq:entropicclocknormalization}
u^a\nabla_a\Theta_{\rm ent}=1 .
\ee
This is a choice of clock units rather than an additional physical restriction.

With this interpretation, self-creation is described by allowing the matter-space density to depend on entropic time,
\be{eq:nABCEntropic}
n_{ABC}=n_{ABC}(N^D,\Theta_{\rm ent}) .
\ee
The particle creation rate then becomes
\be{eq:GammaEntropic}
\Gamma_N
=\nabla_a n^a
=
(\nabla_a\Theta_{\rm ent})\tilde n^a ,
\ee
where
\be{eq:ntildeEntropic}
\tilde n^a
\equiv
\frac1{3!}\epsilon^{abcd}
\frac{\partial N^B}{\partial x^b}
\frac{\partial N^C}{\partial x^c}
\frac{\partial N^D}{\partial x^d}
\frac{\partial n_{BCD}}{\partial\Theta_{\rm ent}} .
\ee
Since \(\tilde n^a\) is constructed from a three-form on the same matter space, it is parallel to \(u^a\). Hence, provided \(u^a\nabla_a\Theta_{\rm ent}\neq0\),
\be{eq:ntildeEntropicGamma}
\tilde n^a
=
\frac{\Gamma_N}{u^b\nabla_b\Theta_{\rm ent}}u^a .
\ee
Under the normalization~\eqref{eq:entropicclocknormalization}, this reduces to Eq.~\eqref{tilde n}.
Thus the particle creation rate measures how the matter-space density changes along the entropic direction.

As in the standard scalar-clock analysis of Sec.~\ref{subsec:standard-scalar-clock}, sliding the labels along the flow probes this creation mode: in a self-creating dissipative system the same displacement changes the value of the entropic clock and hence the amount of matter created or destroyed.
The flow-aligned relabelling therefore probes the thermodynamic arrow of time. The residual proportional to
$
\Gamma_N u^a
$
is the local expression of this irreversible change. It is not a failure of the variational principle, but the manifestation of the physical scalar mode associated with entropic evolution.

The entropic-time interpretation also clarifies the closure of the equations. When particle number is conserved, the scalar equation \(\Gamma_N=0\) closes the number-current sector. When particle creation is present, this equation is replaced by an evolution law for the entropic clock. For example, one may prescribe a constitutive relation
\be{eq:creationlawentropic}
\Gamma_N=\Gamma_N(n,s,\chi,\Theta_{\rm ent},\Gamma_S,\cdots),
\ee
or derive it from a clock-sector action or constraint. In either case, the missing scalar equation is no longer a conservation law. It is the dynamical law that determines how fast the system moves along entropic time.

This point is conceptually important. Particle creation in a dissipative fluid is not merely evolution with respect to coordinate time or proper time. It is evolution along an internally selected thermodynamic direction. The clock field \(\Theta_{\rm ent}\) records this direction locally, and the dependence \(n_{ABC}(N^D,\Theta_{\rm ent})\) encodes how the particle content of each worldline changes as entropy is produced. In this sense, the scalar mode exposed by the loss of flow-aligned relabelling gauge symmetry may be carried by the entropic clock.

\subsection{Towards a local action}
\label{subsec:localaction}
We now comment on what is required for a local action principle for self-creation, in the broader context of variational approaches to irreversible processes~\cite{Ichiyanagi1994}.
The discussion above shows that particle creation can be encoded locally by allowing the matter-space density to depend on a scalar clock~\eqref{eq:nABCTheta}.
In a dissipative setting this clock may be interpreted as the entropic time \(\Theta_{\rm ent}\), but for the moment we keep the notation \(\Theta\) general.

There is an immediate restriction. If \(\Theta\) enters the matter Lagrangian only algebraically through \(n_{ABC}(N^D,\Theta)\), and if it is varied freely as an ordinary scalar field, then its variation gives
\[
\partial_\Theta\Lambda
=\chi_a\tilde n^a ,
\]
where $\tilde n^a$ is given in Eq.~\eqref{eq:ntildeTheta}.
Using Eq.~\eqref{eq:ntildeGamma1} and  $\chi\equiv -u^a\chi_a ,$
one obtains
\[
\partial_\Theta\Lambda
=
-\chi \,\frac{\Gamma_N}{u^c\nabla_c\Theta}.
\]
Therefore, under the weak clock normalization~\eqref{eq:weakclock1}, the free scalar equation \(\partial_\Theta\Lambda=0\) would force
$
\Gamma_N=0
$
unless \(\chi=0\). A freely varied algebraic clock therefore suppresses the very creation process it was introduced to describe.

This shows that a non-trivial local theory of self-creation requires additional structure. The clock must not be a passive algebraic parameter. It must either have its own dynamics, be subject to a constraint, or be coupled to another sector that supplies the entropy-production law. In the entropic-time interpretation, this additional structure is precisely what determines how fast the system moves along the thermodynamic arrow of time.

One possible local implementation is to add a clock-sector term depending on \(\nabla_a\Theta\). Schematically, let
\[
\Lambda_{\rm tot}
=
\Lambda(n^a,s^a;\Theta)+K(\nabla_a\Theta,\Theta,\cdots),
\]
where the dependence of \(\Lambda\) on \(\Theta\) is induced through \(n_{ABC}(N^D,\Theta)\). Variation with respect to \(\Theta\) gives a local clock equation of the form
\be{eq:clockEOM}
\nabla_a\left(\frac{\partial K}{\partial\Theta_{,a}}\right)
- \frac{\partial K}{\partial\Theta}
=  -\chi\,\frac{\Gamma_N}{u^c\nabla_c\Theta}.
\ee
If \(K\) has no explicit \(\Theta\)-dependence and the weak normalization \(u^c\nabla_c\Theta=1\) is imposed, this reduces to
\[
\nabla_a\left(\frac{\partial K}{\partial\Theta_{,a}}\right)
=
-\chi \,\Gamma_N .
\]
Thus the clock equation replaces the conservative scalar equation \(\nabla_a n^a=0\). The creation rate is no longer set to zero; instead it is determined by the dynamics or constraint of the clock.

In the entropic-time interpretation, one may impose the weak normalization by a Lagrange multiplier,
\[
\Lambda_{\rm constr}
=
\lambda\left(u^a\nabla_a\Theta_{\rm ent}-1\right),
\]
or more generally prescribe a monotonic entropy-production relation,
\[
u^a\nabla_a\Theta_{\rm ent}
=\mathcal{F}(\Gamma_S,s,n,\chi,\Theta_{\rm ent},\cdots),
\qquad
\Gamma_S\equiv\nabla_a s^a ,
\]
with \(\mathcal{F}>0\) when irreversible entropy production is present. In this form, the missing scalar equation is not an ad hoc replacement of particle-number conservation. It is the local law that relates particle creation to the irreversible production of entropy.

A diffeomorphism-invariant action must also include the stress-tensor contribution of the clock sector and of any constraint used to define the entropic time. Once these contributions are included, the Noether identity associated with diffeomorphism invariance gives total energy-momentum conservation on shell. In particular, the force balance in the fluid sector need not be conserved separately; exchange with the clock or entropy-production sector is allowed, while the total stress tensor is conserved.
In the entropic-time interpretation, the physical scalar mode exposed for self-creation---the residual proportional to $\Gamma_N u^a$ that, lacking a partner matter space, is not cancelled (Sec.~\ref{subsec:standard-scalar-clock})---is the local thermodynamic direction along which the particle content of a worldline changes, and the clock-sector dynamics constructed above is precisely what governs its evolution.

There is, however, a second possible convention. If one insists on preserving the flow-aligned relabelling as a gauge symmetry even when \(\Gamma_N\neq0\), one may prescribe the clock variation with the opposite sign,
\be{eq:clockReadingOption}
\delta\Theta=+\xi^a\nabla_a\Theta .
\ee
Equivalently, this is the \(\sigma=+1\) choice in
\[
\delta\Theta=\sigma \,\xi^a\nabla_a\Theta .
\]
Then the clock contribution to the relabelling variation changes sign, and the residual becomes
\[
(\delta-\bar\delta)n^a=(-1+\sigma)G\Gamma_Nu^a .
\]
For \(\sigma=+1\), it vanishes identically. This prescription therefore restores the usual gauge interpretation of the flow-aligned relabelling.

This option should be distinguished from the standard scalar-clock interpretation. The standard choice treats \(\Theta\) as an ordinary scalar under the Lagrangian displacement~\eqref{eq:DeltaThetaZero}. With this choice, the non-zero residual is physical and represents the creation mode. The alternative sign choice should instead be regarded as a clock-reading prescription: it reproduces the local endpoint contribution to proper time along the flow and preserves the gauge interpretation of the relabelling, but a first-principles local action realizing this non-standard transformation law remains to be constructed 
(the proper-time functional of Sec.~\ref{sec:propertime} realises $\sigma =+1$, but non-locally).
In the present work we mainly adopt the standard scalar-clock, or entropic-time, interpretation.

\section{Inter-species conversion in the Andersson--Comer construction}
\label{sec:AC}

In the local scalar-clock description of Sec.~\ref{sec:self}, the timelike residual proportional to \(\Gamma_N u^a\) had to be carried by the clock because no partner sector was available. 
We now present a different local realization, in which a second matter space is available and the same residual is absorbed at the level of the coupled force balance: inter-species conversion in the multi-fluid construction of Andersson and Comer~\cite{Andersson:2013jga}; see also Ref.~\cite{AnderssonNew} for a review.
Andersson and Comer showed that resistivity and related dissipative effects can be incorporated by allowing the matter-space forms of different constituents to depend on one another.
This dependence represents the fact that dissipation and particle production arise from interactions between different constituents.

We denote the two matter spaces by the capital labels \(N\) and \(S\).
Their coordinates are \(N^A\) and \(S^A\), and the corresponding matter-space three-forms are denoted by \(n_{ABC}\) and \(s_{ABC}\).
When needed, the same labels \(N\) and \(S\) will also be attached to spacetime quantities, for example to the creation rates \(\Gamma_N\), \(\Gamma_S\), and to the Lagrangian displacements \(\xi_N^a\), \(\xi_S^a\).
For simplicity we discuss two matter spaces only; the extension to more constituents is straightforward.

Let the three-form \(\bm n\) associated with the \(N\)-matter space depend also on the coordinates of the \(S\)-matter space:
\be{int matter}
\bm n
=n_{ABC}(N^D,S^E)\,
dN^A\wedge dN^B\wedge dN^C .
\ee
Such multi-fluid couplings are closely related to entrainment: the conjugate momentum of a constituent need not be aligned with its velocity, a feature important in superfluid and heat-conducting relativistic systems~\cite{Andreev1975,Alpar1984,Andersson11,Monsalvo2011,AnderssonNew}.

Because \(n_{ABC}\) depends on the other matter-space coordinates, the pull-back of \(\bm n\) is not closed in general.
The corresponding number current therefore has a non-vanishing creation rate,
\be{Gamma neq 0}
\Gamma_N
\equiv
\nabla_a n^a
=\sum_{S\neq N}
\frac1{3!}\epsilon^{abcd}
\frac{\partial S^A}{\partial x^a}
\frac{\partial N^B}{\partial x^b}
\frac{\partial N^C}{\partial x^c}
\frac{\partial N^D}{\partial x^d}
\left(
\frac{\partial n_{BCD}}{\partial S^A}
\right) .
\ee
Thus \(\Gamma_N\) measures the dependence of the \(N\)-matter-space density on the coordinates of the other matter spaces.
The creation rate \(\Gamma_S\) is defined in the same way.

We now introduce the Lagrangian displacement \(\xi_N^a\) associated with the \(N\)-flow.
With
\[
\Delta_N\equiv \delta+\Lie_{\xi_N},
\]
the matter-space labels satisfy
\be{D N}
\Delta_N N^A = \delta N^A + \Lie_{\xi_N} N^A =0, 
\ee
because \(N^A\) is comoving with the \(N\)-fluid elements.
The Eulerian variation of the pulled-back three-form is then
\be{d n BCD}
\delta n_{bcd}
=-\Lie_{\xi_N}n_{bcd}
+
\frac{\partial N^B}{\partial x^{[b}}
\frac{\partial N^C}{\partial x^c}
\frac{\partial N^D}{\partial x^{d]}}
\Delta_N n_{BCD},
\ee
where
\be{D n}
\Delta_N n_{BCD}
=\sum_S
\frac{\partial n_{BCD}}{\partial S^E}
\left(
\xi_N^a-\xi_S^a
\right)
\frac{\partial S^E}{\partial x^a}.
\ee

Starting from Eq.~\eqref{d n BCD}, one obtains~\cite{AnderssonNew} 
\bea
\chi_a \delta n^a 
&=& \chi_a \left(n^b \nabla_b \xi^a_N - \xi_N^b \nabla_b n^a -n^a \nabla_b \xi_N^b 
	- \frac12 n^a g^{bc}\delta g_{bc} \right) 
	 - \sum_{S\neq N} R_a^{NS} (\xi^a_S - \xi^a_N) . 
	 \label{chi dn}
\eea
The terms inside the parentheses are the conservative pull-back contribution~\eqref{d na}.
The additional term is defined by
\be{R NS}
R_a^{NS}
\equiv
\frac1{3!}
\chi_N^{BCD}
\frac{\partial n_{BCD}}{\partial S^A}
\frac{\partial S^A}{\partial x^a}.
\ee
By construction, $s^aR_a^{NS}=0,$ and $n^aR_a^{SN}=0.$

The variation of the matter Lagrangian density \(\Lambda(n^a,s^a)\) takes the form, up to total derivatives,
\be{d Lambda}
\delta(\sqrt{-g}\Lambda)
=-\sqrt{-g}
\left\{
\left(
f_a^N+\chi_a\Gamma_N-R_a^N
\right)\xi_N^a
+
\left(
f_a^S+\Theta_a\Gamma_S-R_a^S
\right)\xi_S^a
-\frac12T^{ab}\delta g_{ab}
\right\},
\ee
where
\[
T^{ab}
=\Psi g^{ab}
+
n^a\chi^b+s^a\Theta^b,
\qquad
\Psi
=\Lambda-\chi_an^a-s^a\Theta_a,
\]
and
\be{R x}
f_a^N
\equiv
2n^b\nabla_{[b}\chi_{a]},
\qquad
f_a^S
\equiv
2s^b\nabla_{[b}\Theta_{a]},
\qquad
R_a^N
\equiv
R_a^{SN}-R_a^{NS}
=-R_a^S .
\ee
The equations of motion are therefore
\be{eom:diss}
f_a^N+\Gamma_N\chi_a
=R_a^N,
\qquad
f_a^S+\Gamma_S\Theta_a
=R_a^S .
\ee
Independently of the equations of motion, the definitions of \(\Gamma_N\) and \(R_a^{NS}\), together with the identity~\eqref{chi identity}, imply
\be{u RNS}
u^aR_a^{NS}
=\chi\Gamma_N,
\qquad
n^aR_a^{NS}
=n\chi\Gamma_N .
\ee
The total energy-momentum conservation law follows on shell: 
\[
\nabla_bT^b{}_{a}
= f_a^N+f_a^S+\chi_a\Gamma_N+\Theta_a\Gamma_S
= R_a^N+R_a^S
=0.
\]
The results up to this point are the standard Andersson--Comer multi-fluid force balance.

We next make explicit the corresponding Eulerian variation of \(n^a\).
First, we need a useful identity for \(\chi^{BCD}\).
Define
\[
V^a{}_{BCD}
\equiv
\epsilon^{abcd}
\partial_bN^B\partial_cN^C\partial_dN^D .
\]
This vector is orthogonal to all gradients \(\partial_aN^A\), and hence must be parallel to \(n^a\).
The proportionality factor is fixed by the definition of the conjugate matter-space three-form \(\chi^{BCD}\), giving
\be{chi identity}
\epsilon^{abcd}
\frac{\partial N^B}{\partial x^b}
\frac{\partial N^C}{\partial x^c}
\frac{\partial N^D}{\partial x^d}
=\frac{n^a}{n}\frac{\chi^{BCD}}{\chi}.
\ee
Equivalently,
\be{chi BCD}
\chi^{BCD}
=\chi u_a\epsilon^{abcd}
\frac{\partial N^B}{\partial x^b}
\frac{\partial N^C}{\partial x^c}
\frac{\partial N^D}{\partial x^d}.
\ee
This identity reflects the uniqueness, up to scale, of a totally antisymmetric three-form on the three-dimensional matter space.

From equation~\eqref{d n BCD}, using Eq.~\eqref{chi BCD}, the Eulerian variation of the number current becomes
\bea
\delta n^a
&=&
-\Lie_{\xi_N}n^a
-n^a
\left(
\nabla_b\xi_N^b
+
\frac12 g^{bc}\delta g_{bc}
\right)
-\zeta^a,
\label{delta n zeta}
\\
\zeta^a
&\equiv&
\frac{n^a}{n\chi}
\sum_{S\neq N}
(\xi_N-\xi_S)^eR_e^{NS}.
\nn
\eea
The first two terms are the same as in the conservative pull-back variation, while \(\zeta^a\) is the additional contribution generated by the dependence on the other matter-space coordinates.
Contracting Eq.~\eqref{delta n zeta} with \(\chi_a\) reproduces Eq.~\eqref{chi dn}, as required.

\paragraph{Current-level residual versus action-level degeneracy}---
We can now revisit the relabelling argument.
Consider two variations generated by
\[
\bar\xi_N^a=\xi_N^a-G_N^a,
\qquad
\bar\xi_S^a=\xi_S^a-G_S^a .
\]
The difference between the two variations of the \(N\)-current is
\bea
(\delta-\bar\delta)n^a
&=&
2\nabla_b(n^{[b}G_N^{a]})
-G_N^a\Gamma_N
+
\frac{n^a}{n\chi}
\sum_{S\neq N}
R_e^{NS}
(-G_N^e+G_S^e).
\label{d n 2}
\eea
This is the corrected form of the conservative relabelling formula in the presence of matter-space interactions.

Now take the relabelling vectors to be aligned with the corresponding flows,
\[
G_N^a=G_Nu^a,
\qquad
G_S^a=G_Sv^a,
\]
where \(u^a=n^a/n\) and \(v^a=s^a/s\).
The antisymmetric term in Eq.~\eqref{d n 2} vanishes because \(G_N^a\) is parallel to \(n^a\), and the \(G_S^a\) term vanishes because \(s^aR_a^{NS}=0\).
Using Eq.~\eqref{u RNS}, we therefore find 
\be{current residual AC}
(\delta-\bar\delta)n^a
= -G_N\Gamma_Nu^a
- \frac{n^a}{n\chi}G_Nu^eR_e^{NS}
=-2G_N\Gamma_Nu^a \neq 0.
\ee
Thus the Eulerian current variation itself is not invariant under a flow-aligned relabelling when \(\Gamma_N\neq0\).
The non-conservation of the current leaves a purely timelike residual proportional to the creation rate.

Equation~\eqref{current residual AC} does not mean that the Andersson--Comer construction is inconsistent with a variational principle.
The cancellation occurs at the level of the action variation.
Indeed, the coefficient of \(\xi_N^a\) in Eq.~\eqref{d Lambda} is
\[
\mathcal E_a^N
\equiv
f_a^N+\chi_a\Gamma_N-R_a^N .
\]
Its contraction with the number current vanishes identically:
\bea
n^a\mathcal E_a^N
=
n^af_a^N+n^a\chi_a\Gamma_N-n^aR_a^N
=
0 - n\chi\Gamma_N
 -(-n\chi\Gamma_N)
= 0 .
\label{longitudinal null AC}
\eea
Here we used \(n^af_a^N=0\), \(n^a\chi_a=-n\chi\), and
$
n^aR_a^N
= n^a(R_a^{SN}-R_a^{NS})
 =-n^aR_a^{NS}
=-n\chi\Gamma_N .
$

Therefore, in the Andersson--Comer multi-fluid construction, the extra matter-space dependence does not make the Eulerian current variation invariant under a flow-aligned relabelling.
Rather, it modifies the variational force balance so that the flow-aligned part of the displacement drops out of the action variation.
This cancellation is off shell: it follows from the definitions of the interaction force and the creation rate, not from imposing \(\mathcal E_a^N=0\).
The apparent relabelling obstruction is removed at the level of the action and the equations of motion, not at the level of the current variation itself.

This distinction is important.
Inter-species conversion does not restore the conservative current-level gauge symmetry.
Instead, the timelike residual generated by \(\Gamma_N\neq0\) is absorbed by the partner-sector interaction force \(R_a^N\).
Thus systems with \(\nabla_an^a\neq0\) can still be described within the pull-back variational framework, provided the creation of one species is represented as conversion into, or from, another matter-space sector.

\paragraph{Comparison with the standard clock result}---
Equation~\eqref{current residual AC} has the same timelike structure as the residual found for the standard scalar clock in Sec.~\ref{subsec:standard-scalar-clock}: in both cases, the Eulerian current variation is not invariant under a flow-aligned relabelling when \(\Gamma_N\neq0\). 
The difference is where this residual is assigned. In the single-flow clock description, no partner matter space is available, so the residual is carried by the physical scalar clock mode. 
In the Andersson--Comer construction, the residual remains at the current level but is absorbed at the level of the coupled action variation through the interaction force \(R_a^N\). 

\section{A worked example: particle creation on an FLRW background}
\label{sec:cosmol}

The thermodynamic interpretation of matter creation in cosmology follows the
open-system treatment of Refs.~\cite{Prigogine1988,Calvao1992,Lima1996}.
As an explicit realization in which \(a^b=0\) while \(\Gamma_N\neq0\), we consider a
spatially flat Friedmann--Lema\^itre--Robertson--Walker spacetime,
\be{eq:flrw}
ds^2=-dt^2+a^2(t)\,\delta_{ij}\,\d x^i\d x^j ,
\ee
with a comoving number current \(n^a=n(t)u^a\), \(u^a=(\partial_t)^a\). Spatial
homogeneity forbids any spatial pressure gradient, so the comoving congruence is
geodesic,
$
u^a\nabla_a u^b=0 .
$
With \(H\equiv\dot a/a\), the expansion is
$
\theta=\nabla_a u^a=3H ,
$
and the number current has
\be{eq:GammaFLRW}
\Gamma_N\equiv\nabla_a n^a=\dot n+3Hn .
\ee
Choosing a creation rate per particle 
\be{Gamma}
\Gamma_N= n \Gamma, \qquad \Gamma(t)=3\beta H(t), 
\qquad 0<\beta<1,
\ee
gives
$
\dot n+3Hn=3\beta Hn ,
$
and hence
\be{eq:ndilution}
n(t)=n_0\,a^{-3(1-\beta)}(t) .
\ee
The dilution law~\eqref{eq:ndilution} is the standard one for gravitationally
induced creation of cold dark matter~\cite{Prigogine1988,Prigogine1989,Lima2010},
and the per-particle rate \(\Gamma=3\beta H\)~\cite{Lima1996} corresponds to an
effective equation of state \(w_{\rm eff}=-\beta\).

\subsection{Proper-time realization}
\label{subsec:flrw-propertime}
In the pull-back language, the above history may be represented by the
matter-space three-form
\be{eq:nABCflrw}
n_{ABC}(N^D,\tau)=\mathcal{N}_0\,a^{3\beta}(\tau)\,\epsilon_{ABC},
\ee
with \(N^A=x^A\) the comoving labels. Then \(n=\mathcal{N}/a^3\), where
$
\mathcal{N}\equiv \mathcal{N}_0 a^{3\beta},
$
and Eq.~\eqref{eq:ndilution} is reproduced. Moreover,
\[
\nabla_a n^a=3\beta Hn\neq0 .
\]
From
$\partial_\tau n_{BCD}=3\beta H\,n_{BCD}$ one finds
\be{eq:ntilde-check}
\tilde n^a=\frac1{3!}\epsilon^{abcd}\frac{\partial N^B}{\partial x^b}
\frac{\partial N^C}{\partial x^c}\frac{\partial N^D}{\partial x^d}
\frac{\partial n_{BCD}}{\partial\tau}=3\beta H\,n^a=\Gamma_N\,u^a ,
\ee
so the identity \(\tilde n^a=\Gamma_Nu^a\), used in
Sec.~\ref{sec:propertime}, holds explicitly.

Along the geodesic comoving flow, proper time coincides with cosmic time, \(\tau=t\), and
\be{eq:cosmicclock}
\nabla_a t=-u_a .
\ee
Thus cosmic time is a global scalar clock. Its existence is guaranteed by the
fact that the comoving congruence is geodesic and irrotational. The history term
in the proper-time variation,
$
\int^\tau (a_e\xi^e)\,\d\tau' ,
$
therefore vanishes identically. The Eulerian variation~\eqref{d n} reduces to the
local form
\be{eq:dn-flrw}
\delta n^a
=
 -\Lie_\xi n^a
- n^a
\left(
\nabla_b\xi^b
+
\xi^b n_b\frac{\Gamma_N}{n^2}
+
\half g^{bc}\delta g_{bc}
\right).
\ee
This removes the proper-time history obstruction on the FLRW background.

This realization is the gauge-preserving case discussed in
Sec.~\ref{sec:propertime}. Since the clock is the proper-time functional and the
history term vanishes, the endpoint variation cancels the timelike creation
residual. The flow-aligned relabelling therefore remains a current-level gauge
symmetry even though \(\Gamma_N\neq0\).

The same background also admits an entropic reading. In a homogeneous dissipative
cosmology the entropy production rate is a scalar function of \(t\), and the
thermodynamic arrow of time is aligned with the cosmological expansion. One may
therefore identify, up to a monotonic reparametrization,
\[
\Theta_{\rm ent}=\Theta_{\rm ent}(t),
\qquad
\dot\Theta_{\rm ent}>0 .
\]
The weak clock normalization corresponds to the convenient choice
\(\Theta_{\rm ent}=t\). In this sense, the proper-time clock, cosmic time, and a
local entropic clock can all be chosen to coincide on the FLRW background.

\subsection{Local scalar-clock realization}
\label{subsec:flrw-local-clock}
We now describe the same cosmological history using an independent local scalar
clock. The background geometry, comoving flow, and density history are kept
fixed, but the clock is no longer interpreted as the proper-time functional of
the deformed worldline. Instead, we take
\be{eq:nABCflrwB}
n_{ABC}(N^D,\Theta)=\mathcal{N}_0\,a^{3\beta}(\Theta)\,\epsilon_{ABC},
\ee
where \(\Theta\) is an ordinary scalar field. On the comoving flow the weak clock condition~\eqref{eq:weakclock1},
$ u^a\nabla_a\Theta=1 $ is solved by \(\Theta=t\). 
Since the congruence is geodesic and irrotational,
this is also a strong clock. The background number density and creation rate are therefore unchanged:
$n(t)=n_0\,a^{-3(1-\beta)}$ and $\Gamma_N=3\beta Hn .$
The proper-time and scalar-clock realizations thus describe the same FLRW
history.

They differ in the variational status of the flow-aligned relabelling. For an ordinary scalar clock~\eqref{eq:DeltaThetaZero}, the residual~\eqref{eq:residualStandard} does not vanish:
\be{eq:residualFLRW}
(\delta-\bar\delta)n^a
= -2G\Gamma_N u^a
= -6\beta H\,n\,G\,u^a \neq 0 .
\ee
In the proper-time realization, the same flow-aligned slide is a gauge direction.
In the local scalar-clock realization, it is physical: it changes the scalar
clock reading and hence the amount of matter created along the worldline. The
residual~\eqref{eq:residualFLRW} is therefore the FLRW manifestation of the
scalar creation mode discussed in Sec.~\ref{sec:self}.

Because this flow-aligned direction is physical, the scalar equation that was
\(\Gamma_N=0\) in the conservative theory must be replaced by a clock-sector
equation or constraint. Two simple local closures illustrate the possibilities.

First, one may impose the weak clock condition by a Lagrange multiplier,
\[
\Lambda_{\rm constr}
=\lambda\left(u^a\nabla_a\Theta-1\right).
\]
Variation with respect to \(\lambda\) enforces \(u^a\nabla_a\Theta=1\). Variation
with respect to \(\Theta\), using
$
\partial_\Theta\Lambda=-\chi\Gamma_N
$
under the weak normalization, gives
\be{eq:lambdaEOM}
\nabla_a(\lambda u^a)
=-\chi\Gamma_N .
\ee
On the FLRW background this becomes
\[
\dot\lambda+3H\lambda=-\chi\Gamma_N .
\]
Thus, once a creation law such as \(\Gamma_N=3\beta Hn\) is prescribed, the
multiplier \(\lambda(t)\) carries the scalar closure associated with the clock
sector.

Second, one may give the clock its own dynamics. For example, adding
\[
K=-\frac12\nabla_a\Theta\nabla^a\Theta
\]
leads to the sourced clock equation
\be{eq:boxTheta}
\Box\Theta
=\chi \,\frac{\Gamma_N}{u^c\nabla_c\Theta}.
\ee
For a homogeneous clock on FLRW, this is
\[
\ddot\Theta+3H\dot\Theta
=-\chi \,\frac{\Gamma_N}{\dot\Theta}.
\]
In this case the creation history is determined by the clock dynamics together
with the matter coupling and the \(\Theta\)-dependence of the matter-space
density.

The contrast between the two realizations is therefore sharp. The same FLRW
density history can be represented either by a proper-time clock, for which the
flow-aligned relabelling remains gauge, or by a standard local scalar clock, for
which the same relabelling is a physical scalar creation mode. On this background
the clocks may be numerically identified with cosmic time, but their variational
status is different.

The distinction between the two realizations is invisible at the homogeneous
background level.  Both clocks can be identified with cosmic time on FLRW, and
both reproduce the same dilution law~\eqref{eq:ndilution}.  Their
difference becomes physical at the perturbative level.  In the proper-time
realization, the creation clock is tied to the perturbed fluid worldline and to
the metric, so the perturbation of the creation rate is geometrically
constrained.  In the local scalar-clock realization, by contrast, the clock has
its own scalar perturbation \(\delta\Theta\), whose dynamics or constraint
determines the perturbation of the creation rate.  The two descriptions may
therefore lead to different creation-pressure perturbations, effective sound
speeds, entropy perturbations, and structure-growth histories, even when their
background expansion histories coincide.

\subsection{Perturbative distinction between the two clocks}
\label{subsec:flrw-perturb-clock}

The proper-time and local scalar-clock realizations describe the same homogeneous FLRW history, but they need not describe the same perturbation theory.  
This can be seen already at the level of the creation-pressure perturbation.  
We briefly spell out the distinction, since it gives an observable meaning to the different variational status of the two clocks.
We do not attempt here to solve the full clock-sector dynamics.  Instead, we use a minimal phenomenological closure for \(\Gamma_\Theta\) in order to display the perturbative degree of freedom carried by the local clock.

Consider the background creation law~\eqref{Gamma}, which gives 
$n \propto a^{-3(1-\beta)} $ as in Eq.~\eqref{eq:ndilution}.
Assuming adiabatic particle creation in the usual sense that the specific entropy per particle is conserved, the non-conservative energy balance can be rewritten as a conservative balance with an effective creation pressure.
For a fluid component undergoing adiabatic particle creation, the usual creation pressure may be written in the covariant form
\begin{equation} \label{pc}
        p_c = - \frac{(\rho+p)\Gamma }{\theta},
        \qquad
        \theta \equiv \nabla_a u^a .
\end{equation}
Here \(p_c\) is not a microscopic kinetic pressure.  
It is the effective creation pressure obtained by rewriting the non-conservative energy balance as a conservative fluid equation.  
We will specialize to cold matter, \(p=0\), below.
It is the effective creation pressure obtained by rewriting the
non-conservative energy balance as a conservative fluid equation.
On the homogeneous background, \(\theta=3H\), and therefore
$p_c = -\beta(\rho+p) .$
For cold matter, \(p=0\), this becomes
\begin{equation}
        p_c = -\beta\rho ,
\end{equation}
so that the effective background equation of state is $w_{\rm eff} = -\beta .$

We now perturb the creation pressure.  From Eq.~\eqref{pc} one obtains
\begin{equation}
        \delta p_c
        =
        -\frac{\Gamma}{\theta}(\delta\rho+\delta p)
        -\frac{\rho+p}{\theta}\delta\Gamma
        +\frac{(\rho+p)\Gamma}{ \theta^2}\delta\theta .
\end{equation}
Using \(\Gamma/\theta=\beta\) on the background, this becomes
\begin{equation}
        \delta p_c
        =
        -\beta(\delta\rho+\delta p)
        -\frac{\rho+p}{ 3H}
        \left(
             \delta\Gamma-\beta\delta\theta
        \right).
        \label{eq:delta-pc-general}
\end{equation}
Thus the perturbative distinction between different clock realizations is
controlled by the combination
\begin{equation}
        \Delta_\Gamma
        \equiv
        \delta\Gamma-\beta\delta\theta .
        \label{eq:DeltaGamma}
\end{equation}

For the proper-time-clock realization, the clock is tied to the deformed
fluid worldline.  The natural covariant completion of the background law~\eqref{Gamma} is therefore
$ \Gamma = \beta\theta .$
Consequently, $
        \delta\Gamma = \beta\delta\theta ,$
and hence
\begin{equation}
        \Delta_\Gamma = 0 , \qquad \mbox{(proper-time clock).}
\end{equation}
Equation~\eqref{eq:delta-pc-general} then reduces to
\begin{equation}
        \delta p_c^{\rm PT}
        =
        -\beta(\delta\rho+\delta p).
        \label{eq:delta-pc-PT-general}
\end{equation}
In particular, for cold matter,
\begin{equation}
        \delta p_c^{\rm PT}
        =
        -\beta\delta\rho .
        \label{eq:delta-pc-PT}
\end{equation}
Since the background pressure is \(p_c=-\beta\rho\), the adiabatic sound speed of the effective component is
$c_{a}^2 \equiv \frac{\dot p_c}{\dot\rho}=  -\beta ,$ and Eq.~\eqref{eq:delta-pc-PT} gives
\begin{equation}
        \delta p_{\rm nad}^{\rm PT}
        \equiv
        \delta p_c^{\rm PT} - c_{a}^2\delta\rho
        =
        0 .
        \label{eq:nonadiabatic-PT}
\end{equation}
Thus the proper-time realization has no independent intrinsic entropy
perturbation in this minimal completion.  Its creation-pressure
perturbation is fixed by the density perturbation.

The local scalar-clock realization is different.  Let
\begin{equation}
        \Theta = t+\delta\Theta ,
\end{equation}
and write the matter-space density as in Eq.~\eqref{eq:nABCflrwB}.
The perturbation \(\delta\Theta\) is not gauge invariant by itself.  This does not affect the physical conclusion, because the clock distinction is encoded in the gauge-invariant mismatch between the creation-rate perturbation and the expansion perturbation.  
Under a first-order time-slicing change \(t\to t+T\), scalar
perturbations transform as
\[
        \delta\Gamma\to \delta\Gamma-\dot\Gamma\,T,
        \qquad
        \delta\theta\to \delta\theta-\dot\theta\,T .
\]
Since the background relation is \(\Gamma=\beta\theta\), with constant \(\beta\), one has \(\dot\Gamma=\beta\dot\theta\).  Therefore $\Delta_\Gamma$ in Eq.~\eqref{eq:DeltaGamma}
is gauge invariant at first order.
In the proper-time realization \(\Delta_\Gamma=0\).  In the local scalar-clock realization, \(\Delta_\Gamma\) is generally non-zero, because the clock perturbation contributes to \(\delta\Gamma_\Theta\) independently of the perturbed
expansion.
The local clock need not be tied to the perturbed expansion scalar.  A minimal local completion of the same background law is
\begin{equation}
        \Gamma_\Theta
        =
        3\beta H(\Theta)\,C ,
        \qquad
        C \equiv u^a\nabla_a\Theta .
        \label{eq:Gamma-local-clock}
\end{equation}
Here \(H(\Theta)\) denotes the background Hubble function evaluated at the clock reading; it is used only as a minimal FLRW closure, not as a general covariant scalar prescription.
In this perturbative diagnostic we impose the weak clock normalization only on the homogeneous background.  
Imposing \(C=1\) also at first order would be a particular clock-sector constraint, setting \(\delta C=0\).
Here we keep \(\delta C\) explicit in order to display the scalar
clock-rate perturbation carried by the local clock.
On the background, \(C=1\) and \(\Theta=t\), so $ \Gamma_\Theta = 3\beta H .$
However, at first order,
\begin{equation}
        \delta\Gamma_\Theta
        =
        3\beta
        \left[
             \dot H\,\delta\Theta
             + H\,\delta C
        \right].
        \label{eq:delta-Gamma-local-general}
\end{equation}
In Newtonian gauge,
\begin{equation}
        ds^2
        =
        -(1+2\Phi)dt^2
        +a^2(t)(1-2\Psi)\delta_{ij}dx^i dx^j ,
\end{equation}
one has, for \(\Theta=t+\delta\Theta\),
\begin{equation}
        \delta C
        =
        \delta\dot\Theta-\Phi .
\end{equation}
Therefore
\begin{equation}
        \delta\Gamma_\Theta
        =
        3\beta
        \left[
             \dot H\,\delta\Theta
             + H(\delta\dot\Theta-\Phi)
        \right], \qquad \mbox{(local-scalar clock).}
        \label{eq:delta-Gamma-local}
\end{equation}
Substitution into Eq.~\eqref{eq:delta-pc-general} gives, for cold matter,
$ \delta p_c^\Theta =
        -\beta\delta\rho
        -\frac{\rho}{ 3H}
        \left(
             \delta\Gamma_\Theta-\beta\delta\theta
        \right).$
Equivalently,
\begin{equation}
        \delta p_c^\Theta
        =
        -\beta\delta\rho
        -\rho\beta
        \left[
             \frac{\dot H}{ H}\delta\Theta
             +\delta\dot\Theta-\Phi
             -\frac{1}{ 3H}\delta\theta
        \right].
        \label{eq:delta-pc-local-expanded}
\end{equation}
The bracketed mismatch term in Eq.~\eqref{eq:delta-pc-local-expanded} is absent in the
proper-time realization.  It is the perturbative imprint of the independent
clock mode.  The corresponding non-adiabatic pressure perturbation is
\begin{equation}
        \delta p_{\rm nad}^{\Theta}
        =
        \delta p_c^\Theta - c_{a}^2\delta\rho
        =
        -\frac{\rho}{3H} \Delta_\Gamma.
        \label{eq:nonadiabatic-local}
\end{equation}
Thus the physical distinction between the two clock prescriptions is quantified not by the gauge-dependent variable \(\delta\Theta\) itself, but by the gauge-invariant quantity \(\Delta_\Gamma\).
Consequently, the local scalar clock generically carries an entropy perturbation, unless its perturbation is constrained so that $\Delta_\Gamma =0$.
This is precisely the condition that collapses the local scalar-clock
description back to the proper-time or expansion-clock behavior at the
linear level.

The distinction may also be expressed in terms of an effective rest-frame
sound speed.  For cold matter one may write schematically
\begin{equation}
        c_{\rm eff}^2
        =
       \left( \frac{\delta p_c}{\delta\rho}\right)_{\rm rf}
\end{equation}
in the appropriate rest frame.  The proper-time realization gives
\begin{equation}
        c_{{\rm eff}, {\rm PT}}^2
        =
        -\beta ,
\end{equation}
whereas the local scalar-clock realization gives
\begin{equation}
        c_{{\rm eff}, \Theta}^2
         =
        -\beta
        -\frac{\rho}{3H\,\delta\rho_{\rm rf}}
        \Delta_{\Gamma,{\rm rf}} .
        \label{eq:ceff-local}
\end{equation}
The second term depends on the clock-sector dynamics or constraint.  Hence
the local scalar-clock realization is not merely a different parametrization
of the same background particle creation history.  It contains an additional
scalar perturbation, which can modify the creation-pressure perturbation,
the non-adiabatic pressure, and the effective sound speed.

Within the effective one-fluid rewriting used in this subsection, the same gauge-invariant mismatch also affects the curvature perturbation.  
On large scales, neglecting anisotropic stress, gradient terms, and any separately resolved clock-sector stress-energy, the curvature perturbation on uniform-density hypersurfaces obeys
\begin{equation}
        \dot\zeta
        =
        -\frac{H} {\rho+p_{\rm eff}}\,
        \delta p_{\rm nad}
        +\cdots .
\end{equation}
For the cold creation model considered here,
\(p_{\rm eff}=p_c=-\beta\rho\), so that
\(\rho+p_{\rm eff}=(1-\beta)\rho\).  Therefore
\begin{equation}
        \dot\zeta
        =
        \frac{1}{ 3(1-\beta)}
        \Delta_\Gamma
        +\cdots .
\end{equation}
The proper-time realization has \(\Delta_\Gamma=0\) and therefore does not
source \(\zeta\) through this channel.  A local scalar clock, by contrast,
generically has \(\Delta_\Gamma\neq0\), and can source curvature
perturbations through its non-adiabatic creation pressure.

This provides the perturbative meaning of the distinction between the two
clock prescriptions.  On the homogeneous FLRW background, proper time,
cosmic time, and a local entropic clock may all be chosen to coincide.  At
linear order, however, the proper-time clock is locked to the perturbed
worldline and expansion, giving \(\Delta_\Gamma=0\), while a standard local
scalar clock generically gives \(\Delta_\Gamma\neq0\).  The gauge-invariant
mismatch \(\Delta_\Gamma\), rather than the gauge-dependent
\(\delta\Theta\) itself, is the physical scalar quantity that distinguishes
the two descriptions.  It sources non-adiabatic creation pressure, modifies
the effective rest-frame sound speed, and can source the curvature
perturbation on large scales.
Thus cosmological perturbations can distinguish the gauge-preserving
proper-time realization from the residual-bearing local scalar-clock
realization.

\section{Summary and discussion} \label{sec:summary}

We have revisited how particle creation modifies the flow-aligned relabelling structure of the pull-back variational formulation of relativistic fluids.
 In the standard conservative construction, the Eulerian variation of the number current is compatible with flow-aligned relabellings only when
\[
\Gamma_N\equiv \nabla_a n^a=0 .
\]
This condition is often interpreted as a variational reason for imposing particle-number conservation. 
The main point of this work is that particle creation does not simply invalidate the pull-back variational framework; rather, it reorganizes the flow-aligned relabelling structure.
Depending on how the creation clock or the interacting matter-space sector is introduced, the timelike relabelling residual is either cancelled, carried by a physical scalar clock mode, or absorbed into a coupled force balance.

The results can be organized by asking whether the flow-aligned relabelling remains a current-level gauge symmetry. There are two qualitatively different possibilities. In the first, the relabelling symmetry is preserved. This is realized when the creation clock is identified with the proper-time functional of the deformed worldline. In that case, the endpoint variation of the proper time cancels the timelike creation residual in the current variation. Thus the flow-aligned relabelling can still be treated as a gauge direction even when the number current is not conserved.
The price of this gauge-preserving prescription is that the proper-time clock is not an ordinary local scalar field. Its variation contains the history term
\[
\int^\tau (a_b\xi^b)\,\d\tau' ,
\]
where \(a^b=u^c\nabla_cu^b\) is the acceleration of the flow. Hence the non-locality is not caused by particle creation itself, but by identifying the creation clock with the proper-time functional of the deformed path. On geodesic flows, \(a^b=0\), this history term vanishes. This is why homogeneous FLRW cosmological backgrounds provide a particularly clean realization of the proper-time-clock prescription.

The second possibility is the local description. 
Here the matter-space density is allowed to depend either on an independent scalar clock or on another matter space. When a clock is present, it is treated as an ordinary local scalar field.
In this case, the flow-aligned relabelling is no longer a current-level gauge symmetry when \(\Gamma_N\neq0\). The current variation contains \emph{a purely timelike residual} proportional to
\[
\Gamma_Nu^a .
\]
This residual expresses the fact that sliding the labels along a worldline changes the particle content assigned to that worldline. The question then becomes how this residual is compensated in the variational principle.

The first local realization is self-creation, a process in which the particle number of a single species changes without being represented as conversion from another matter-space current. In this case there is no partner-sector force \(R_a^N\) that can absorb the timelike residual. The residual must therefore be carried by an additional scalar datum. This is the role of the creation clock. If the matter-space density depends on a standard scalar clock,
\[
n_{ABC}=n_{ABC}(N^D,\Theta),
\]
and \(\Theta\) is varied as an ordinary scalar, \(\Delta\Theta=0\), the variation remains local but the flow-aligned relabelling becomes physical. The formerly redundant scalar direction becomes the degree of freedom that records particle creation along the worldline.

This gives a natural interpretation to the clock in a dissipative system. Entropy production selects a local thermodynamic arrow of time, and the creation clock may be identified with a local entropic time \(\Theta_{\rm ent}\). Then
$
n_{ABC}=n_{ABC}(N^D,\Theta_{\rm ent})
$
means that the particle content attached to a worldline changes along the irreversible direction selected by entropy production. The timelike residual \(\propto \Gamma_Nu^a\) is then the variational imprint of motion along entropic time. In this interpretation, the loss of the conservative flow-aligned relabelling gauge symmetry is not a pathology; it is the appearance of the physical scalar mode associated with irreversible creation.

The second local realization is inter-species conversion in the Andersson--Comer multi-fluid construction. 
Unlike the self-creation cases, inter-species conversion does not require an independent creation clock.  
The conversion information is carried by the dependence of one matter-space density on the coordinates of the partner matter space, and the corresponding timelike residual is absorbed
by the relative force balance.
The Eulerian current variation of a single species is still not invariant under a flow-aligned relabelling; the timelike residual remains at the current level. 
However, the partner matter-space sector supplies an interaction force. If
$
\mathcal E_a^N
\equiv
f_a^N+\chi_a\Gamma_N-R_a^N
$
is the coefficient of the \(N\)-fluid displacement in the action variation, then
\[
n^a\mathcal E_a^N=0 .
\]
Thus the flow-aligned part of the displacement drops out of the coupled action variation. In this sense, the apparent relabelling obstruction is removed at the level of the action and the equations of motion, not at the level of the individual current variation. The residual is absorbed by the relative force balance between the interacting matter-space sectors.

The local clock description also clarifies the equation counting. In the conservative theory, the flow-aligned relabelling is redundant and \(\Gamma_N=0\) supplies the scalar equation for the number-current sector. Once particle creation is allowed, this scalar equation must be replaced. In self-creation, where no partner sector is present, the missing scalar equation must come from the clock sector itself: from a prescribed creation law, a constraint, or an equation of motion derived from an additional clock action. In inter-species conversion, by contrast, the replacement is provided by the coupled force balance between matter spaces. A freely varied algebraic clock would force \(\Gamma_N=0\), while a clock with its own dynamics or constraint can support a non-trivial local creation law.

It is useful to distinguish the local scalar-clock description from the proper-time prescription. A standard scalar clock obeys the usual variational rule $\Delta\Theta=0,$ i.e., $\delta\Theta=-\xi^a\nabla_a\Theta .$
This leads to the residual-bearing local description described above. By contrast, the proper-time functional effectively supplies the opposite endpoint variation under a flow-aligned displacement. Equivalently, one may formally write a non-standard clock variation
$\delta\Theta=+\xi^a\nabla_a\Theta .$ 
This cancels the current-level residual and preserves the flow-aligned relabelling as a gauge direction. However, this is not the standard Eulerian variation of an ordinary scalar field. It should be regarded as a clock-reading prescription associated with the proper-time endpoint. A first-principles local action realizing such a non-standard transformation law remains an open problem.

Finally, the FLRW example illustrates how these distinctions are realized in cosmology.  
In a homogeneous and isotropic background, the comoving congruence is geodesic and irrotational.  
Cosmic time is therefore an orthogonal scalar clock, and one may have particle creation without producing the proper-time history term. 
For example, the creation law \(\Gamma=3\beta H\) gives
\(n\propto a^{-3(1-\beta)}\), the standard dilution law for gravitationally induced creation of cold dark matter.

At the background level, the proper-time realization and the local scalar-clock realization can therefore describe the same cosmological history: proper time, cosmic time, and a local entropic clock may all be chosen to coincide with the unique homogeneous time direction.  
The distinction is nevertheless not only a matter of parametrization.  
With the proper-time clock, the flow-aligned relabelling remains a gauge direction, whereas with a standard local scalar clock the same direction is a physical scalar creation mode.

This difference is hidden by the homogeneous FLRW background but appears already at first order in perturbations.  
In the proper-time realization, the creation-rate perturbation is fixed by the perturbed worldline and expansion, giving the gauge-invariant mismatch \(\Delta_\Gamma=0\).  
A local scalar clock, by contrast, can carry an independent clock-sector perturbation whose gauge-invariant imprint is
\(\Delta_\Gamma\neq0\).  
This mismatch sources non-adiabatic creation pressure, can modify the effective sound speed, and can source curvature perturbations in the effective one-fluid description.  
Thus the homogeneous FLRW background hides the clock distinction, while perturbations reveal that the two realizations define different hydrodynamic theories.

The overall conclusion is that particle creation is compatible with pull-back variational fluid dynamics, but it reorganizes the relabelling structure. A proper-time functional clock preserves the flow-aligned relabelling at the price of a possible history term. A local scalar or multi-fluid description does not preserve the current-level relabelling gauge symmetry; instead it leaves a timelike residual proportional to the creation rate. In single-flow self-creation this residual becomes the physical clock mode, while in inter-species conversion it is absorbed by the partner-sector force balance. This organization of the residual is the central distinction between the different variational descriptions of particle creation.

\begin{acknowledgments} 
This work was supported by the National Research Foundation of Korea (NRF) grants RS-2026-25483539 (H.K.).
\end{acknowledgments}



\end{document}